\documentclass[final,authoryear,5p]{elsarticle}

\usepackage{graphicx}

\usepackage{amssymb}

\usepackage[ps2pdf,%
a4paper=true,%
breaklinks=true,%
colorlinks=true,%
pdfauthor={First Author et al.},%
pdftitle={Template for manuscripts in Advances in Space Research}%
]{hyperref}

\journal{Advances in Space Research}

\begin{document}

\begin{frontmatter}



\title{Anomalies in radiation-collisional kinetics of Rydberg atoms induced by the effects of dynamical chaos and the double Stark resonance}

\author[label1,label3]{N.N.Bezuglov}
\author[label1]{A.N.Klyucharev}
\ead{anklyuch@gmail.com}
\author[label2]{A.A.Mihajlov}
\ead{mihajlov@ipb.ac.rs}
\author[label2]{V. A. Sre{\'c}kovi{\' c}}
\ead{vlada@ipb.ac.rs}
 \address[label1]{Department of physics, Saint-Petersburg University, Ulianovskaya 1, 198504, St.Petersburg, Petrodvorets, Russia}
 \address[label2]{University of Belgrade,Institute of Physics, P. O. Box 57, 11001 Belgrade,Serbia}
 \address[label3]{Laser Centre, University of Latvia, Zellu Str. 8, LV-1002 Riga, Latvia}





\begin{abstract}

Radiative and collisional constants of excited atoms contain the matrix elements of the dipole transitions
and when they are blocked one can expect occurring a number of interesting phenomena in radiation-collisional
kinetics. In recent astrophysical studies of IR emission spectra it was revealed a gap in the radiation emitted
by Rydberg atoms ($RA$) with values of the principal quantum number of $n\approx10$. Under the presence of
external electric fields a rearrangement of $RA$ emission spectra is possible to associate with manifestations
of the Stark effect. The threshold for electric field ionization of $RA$ is $E\approx3\cdot10^{4}$ V/cm for
states with $n>10$. This means that the emission of $RA$ with $n\ge10$ is effectively blocked for such fields.
In the region of lower electric field intensities the double Stark resonance (or F\"{o}rster resonance) becomes
a key player. On this basis it is established the fact that the static magnetic or electric
fields may strongly affect the radiative constants of optical transitions in the vicinity of the F\"{o}ster
resonance resulting, for instance, in an order of magnitude reduction of the intensity in some lines. Then, it is shown in this
work that in the atmospheres of celestial objects lifetimes of comparatively long-lived $RA$ states
and intensities of corresponding radiative transitions can be associated with the effects of dynamic chaos via
collisional ionization. The F\"{o}ster resonance allows us to manipulate the random walk of the
Rydberg electron ($RE$) in the manifold of quantum levels and hence change
the excitation energies of $RA$, which lead to anomalies in the IR spectra.

\end{abstract}


\end{frontmatter}

\parindent=0.5 cm

\section{Introduction}

The aim of this work is to show that in the atmospheres of celestial objects lifetimes of comparatively long-lived states
of the Rydberg atoms ($RA$) and intensities of corresponding radiative transitions can be associated with the
effects of dynamic chaos via collisional ionization.
Inelastic atom-atom collisions in the thermal range of energy (chemi-ionization processes) involving Rydberg atoms ($RA$) and leading to the formation of molecular and atomic ions
\begin{equation}
\label{eq:RA}
RA + A =    \left\{
             \begin{array}{lll}
               A_{2}^{+} + e, \\
                 \\
             \displaystyle{ A^{+} + A + e }
             \end{array}
              \right .
\end{equation}
are traditionally considered in the astrophysical literature as an alternative to elastic $RA + A$ collisions (see, e.g., \citet{kly10}). This is especially true for hydrogen atoms (the solar photosphere) and helium atoms (helium-rich cool white dwarfs stars) \citep{mih11a}. Paper \citet{kly07} deals with the processes of chemi-ionization involving a sodium atom that seems to influence the processes in the atmosphere of Jupiter's satellite Io. The theory and the experiment related to the Rydberg atoms \citep{mih12,oke12,sre13} and especially as an extreme case Rydberg matter with a high degree of ionization and consisting of Rydberg atoms are discussed in literature (see, e. g., \citet{hol06}).

As chemi-ionization processes affect the optical characteristics and profiles of spectral lines emitted in the atmospheres of space objects, they are more and more consistently included in the models of atmospheres of cooler stars and stellar atmospheres \citep{mih11b,gne09}.

It has long been believed that in the absence of collisions (excluding optical transitions) the selectivity of the primary optical transition is preserved. This proposition was questioned in 1988 (see, e.g., \cite{kly07}).

\begin{figure}
\begin{center}
\includegraphics[width=0.8\columnwidth,
height=0.65\columnwidth]{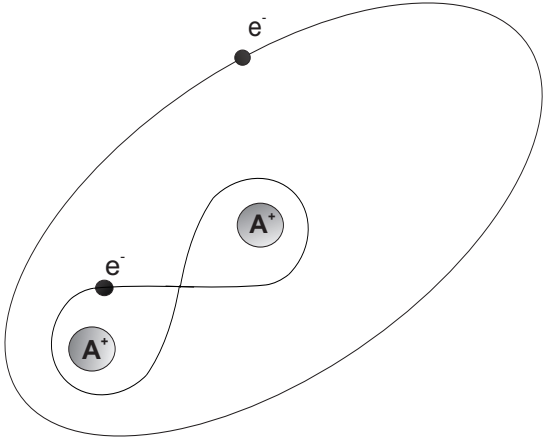}
\end{center}
\caption{Formation of the intermediate quasi-molecular complex in $RA + A$ thermal collisions.}
\label{fig:fig1}
\end{figure}

The information obtained by research examining the above-mentioned processes can be briefly reduced to a few statements:
\begin{enumerate}
  \item	The application of the dipole resonance ionization mechanism is well established within the framework of Fermi model of inelastic collisions $RA + A$ (Fig. \ref{fig:fig1}). This is confirmed by comparing the results of theoretical calculations and experimental results for hydrogen atoms and hydrogen-like alkali atoms with the values of effective quantum numbers $5 \le n \le 30$.
  \item Paper \cite{kly07} drew attention to the need for the chemi-ionization model involving $RA$ to take into account the multiplicity of quasi-crossings of the initial term of the Rydberg molecule ($RA + A$) with a network of terms related to the neighboring states of a selectively excited atom (Fig. \ref{fig:fig2}). As a result of the latter circumstance the application of the classical Landau-Zener theory determining the ionization process (1) for $R \ge R_{c}$ is limited and the connection between the measured rate constant and the initial excited state of $RA$ is questionable.
  \item One of the direct ways to solve this contradiction is associated with the use of the dynamic chaos model in the model of atom-atom collisions \cite{kly10}.
  \item It has been noted that there is an experimentally observed gap in the spectrum of the infrared radiation of white dwarfs. This gap corresponds to optical transitions in $RA$ with $n \approx 10$.  According to \cite{gne09}, this effect can be explained by the following:  taking into account the collision-induced processes of light absorption, the relativistic quantum defect of "vacuum polarization," the influence of strong magnetic fields ($B \ge 10^5$ G), the Stark effect with the electric field intensity $E \ge 10^{6}$Vcm$^{-1}$
\end{enumerate}

It is known that the behavior of a Rydberg electron ($RE$) in the electric field and in the magnetic field is different. The energy of $RE$ interaction with the external electric field of $E$ intensity is
\begin{equation}
\label{eq:We}
W_{E} \approx n^{2}\cdot E
\end{equation}
For $RE$ in a strong magnetic field with induction $B$ the similar expression can be written as:
\begin{equation}
\label{eq:Wb}
W_{B} \approx \frac{1}{8\alpha^{2}n^{4}B^{2}}
\end{equation}
where $W_{B}$ is the energy of $RE$ diamagnetic interaction with the magnetic field, $\alpha$ is the fine structure constant.
Hence:
\begin{equation}
\label{eq:WeWb}
\frac{W_{E}}{W_{B}} \approx n^{-2}\frac{E}{B^{2}}
\end{equation}
Due to the quadratic dependence (\ref{eq:WeWb}) on $n$, the behavior of $RE$ in strong fields is different from the case of atoms in lower excited states. Discrete terms $RA$ under the influence of an external electric field transit into the state of the continuous spectrum, neglecting the tunneling effect at
\begin{equation}
\label{eq:En}
E(n) = \frac{1}{16n^{4}} \, \textrm{a.u.}
\end{equation}
For states with $n = 10 - 12$, this corresponds to $E \le 30$kV$\cdot$cm$^-1$, where "blocking" of optical transitions from upper excited states begins.

The imposition of a magnetic field, as opposed to an electric one, does not lead directly to the ionization of $RA$. When $B \le 2\cdot10^7$ G, it, on the contrary, increases the connection between $RE$ and the atomic core (the case of a "weak" magnetic field). In this case, $n$ remains a "good" quantum number for the excited atom, and the mixing of excited states is much smaller than the energy of $RE$ connection. In a strong magnetic field, when diamagnetic interaction exceeds Coulomb interaction, one can speak about the case of a "free" electron in the external magnetic field. This particularly leads to the observation of Landau resonances in the absorption of light.

The scale of the interparticle interaction in a pseudo-bound complex (cold particles + photon) becomes comparable to the de Broglie wavelength ($T \le 10^{-3}$ K), while the energy transferred in the interparticle collisions is less than the energy transmitted to them by photons. The first experiments that started the above work used cesium $RA$ with $n = 10 \div 12$ in the electric field with the intensity of a few hundred V$\cdot$cm$^{-1}$. Interestingly, it is to these values of the principal quantum number that the maximum of cross-sections of associative ionization ($AI$) of hydrogen and hydrogen-like alkali atoms corresponds. For the sake of completeness of the topic of "blocking" optical transitions in external fields, let us remember that the physical literature has discussed the influence of external fields on the optical properties of the excited atom since 1896 (the Zeeman effect).

\section{Rydberg atoms. Approximation of stochastic dynamics.}
\label{Section RA}

\subsection{Diffusion approach in the collision kinetics problems.}
\label{Section 2RA}

\begin{figure}
\begin{center}
\includegraphics[width=\columnwidth,
height=0.75\columnwidth]{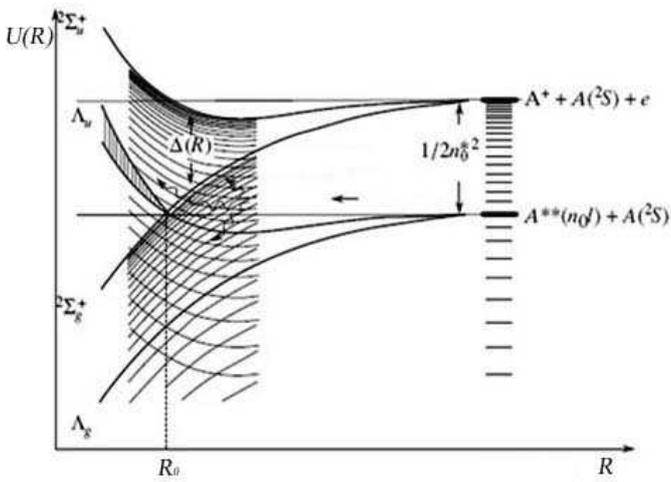}
\end{center}
\caption{Schematic representation of the chemi-ionization processes with $RA$ participation.}
\label{fig:fig2}
\end{figure}

\begin{figure}
\begin{center}
\includegraphics[width=\columnwidth,
height=0.75\columnwidth]{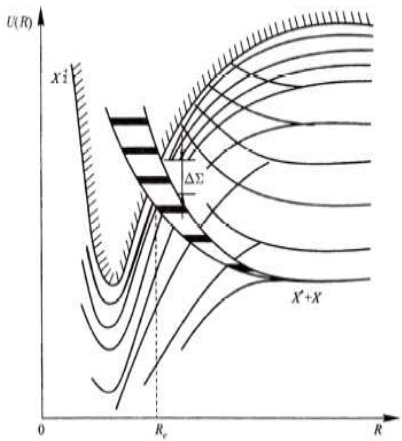}
\end{center}
\caption{Qualitative illustration of the quasicrossing of the covalent and ionic terms under the diffusion approach.}
\label{fig:fig3}
\end{figure}

The possibility of using "diffusion" kinetics in a single atom-atom collision involving Rydberg atoms ($RA$) was first considered in \cite{dev88}. Its authors drew on the fact that it is hardly possible to account for the multiplicity of quasi-crossings of terms of the Rydberg collision complex ($RA + A$) with the terms of the nearest molecular states $A_{2}^{*}$, if, in addition, one takes into account the phenomenon of the Coulomb condensation of terms (Fig. \ref{fig:fig2}). At the same time, the change in the energy of a Rydberg electron ($RE$) in one such avoided crossing at sufficiently large internuclear distances in complex $\Delta\varepsilon$ is the value comparable with the distance between the adjacent energy levels of $RA$ of the order $n^{-3}$, where $n$ is the effective value of the principal quantum number of the excited atom. Thus, the relationship between $\Delta\varepsilon$ and the binding energy of $RE$, $\varepsilon_{0} = 1/(2n^{2})$ is $\Delta\varepsilon/\varepsilon_{0} = n^{-1} \ll 1$. This makes it possible to apply the diffusion approach to chemi-ionization processes in thermal and subthermal collisions. As opposed to the well-known Pitaevskii diffusion model in multiple particle collisions, in the latter case we deal with the diffusion of $RE$ in the web of the terms of energy states of Rydberg complexes in single atom-atom collisions. As a result, the originally single selective term of a quasimolecule is transformed into a conical "bundle" at $\varepsilon_{0} = 1/(2n^2)$ (Fig.\ref{fig:fig3}). When simulating the process leading to ionization (\ref{eq:RA}), let us limit ourselves to the adiabatic approximation of particle motion and the quasi-classical representation of a single trajectory. The terms corresponding to the main terms of ion $A_{2}^{+}$ and molecule $A_{2}^{*}$ are split into two components $\Sigma_{u}$ , $\Sigma_{g}$ and $\Lambda_{u}$, $\Lambda_{g}$ correspondingly (Fig.\ref{fig:fig2}). Hereinafter the atomic system of units is used, unless otherwise stated. The value $R=R_{0}$ corresponds to the beginning of autoionization of the complex at $R < R_{0}$. A more recent paper \citet{bez02} analyzed transient kinetic equations describing the stochastic diffusion of $RE$ in energy space when $R < R_{0}$. It was supposed that in a quasi-monochromatic field emerging in process (\ref{eq:RA}) according to the model of the dipole resonance mechanism, there are internal nonlinear dynamic resonances caused by the coincidence of overtones of $RE$ rotation speed in the Keplerian orbit with the frequency of recharging $\Delta R$. As a result, the $RE$ trajectory motion becomes unstable, and quasi-molecular complex $(RA + A)$, as a consequence of the $RE$ orbit instability with respect to small perturbations, goes into the $K$-system or the dynamic chaos mode. The $RE$ motion becomes a random "walk" on the web of quasi-molecular terms, and the calculation of the $RE$ time distribution function throughout the states with different values of $n$ is reduced to the solution of Kolmogorov-Fokker-Planck equation. Besides, in the problems of this kind one should basically take into account the mixing of states with different values of the orbital quantum number $l$. The key condition for the emergence of global dynamic chaos is overlapping of $n$-adjacent nonlinear resonances of finite width.

\begin{figure}
\begin{center}
\includegraphics[width=0.75\columnwidth,
height=0.65\columnwidth]{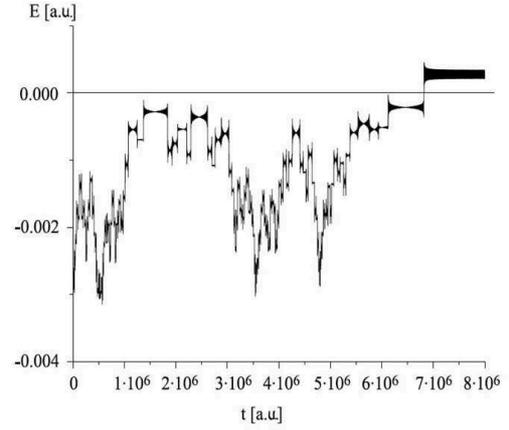}
\end{center}
\caption{Chaotic (diffusion) dynamic of $RE$ energy due to coupling with the resonant external microwave field \citep{gne09}.}
\label{fig:fig4}
\end{figure}

The development of such a situation in a variable microwave electric field, the applicability of the quasi-classical description of the stochastic dynamics of $RE$, and the difference between the occurring ionization process and multiphoton or tunnel ionization, were demonstrated (experimentally and theoretically) in the 1980s (see, e.g., \cite{leo78}). Prohibition of global dynamic chaos mode in atomic (molecular) systems is primarily associated with the dimension of such systems. In the case of a time-varying external disturbance, chaos is possible even for a one-dimensional Hamiltonian system of a hydrogen atom in a microwave electric field.
Average time $\tau_{eff}(n)$ required for an $RE$ to achieve the ionization threshold in the stochastic diffusion mode is:
\begin{equation}
\label{eq:teff}
\tau_{eff}(n) = \frac{\omega_{L}^{4/3}}{0.65\cdot F^{2}}n^{3}\left(\frac{1}{n}-\frac{n_{c}}{2n^{2}}\right)
\end{equation}
where $\omega_{L}$ is the microwave field frequency, $n_{c}$ is the value of the $RA$ principal quantum number $n$ separating the region of chaotic ($n > n_{c}$) and regular ($n < n_{c}$) $RE$ motion and $F$ is the intensity of the microwave field (see Fig.\ref{fig:fig4}). The value of the diffusion coefficient at which $RE$ at the initial time in the state with energy $\varepsilon = -1/(2n^{2})$ reaches the ionization threshold ($n = \infty$), is
\begin{equation}
\label{eq:Dn}
D_{n} = 0.65\cdot F^{2}n^{3}\omega_{L}^{-4/3}
\end{equation}

\section{Specific features of RE stochastic diffusion under the "double Stark resonance" - F\"{o}rster resonance.}
\label{Section 3RA}
Processes associated with the emergence of global chaos for atoms other than hydrogen atoms can be conveniently simulated within the framework of the Sommerfeld model of the excited atom that treats the potential $RE$ motion
\begin{equation}
\label{eq:Ur}
U(r) = -\frac{1}{r}+\frac{\alpha}{2r^2},
\end{equation}
A dimensionless quantity $\langle N \rangle$ can serve as a convenient parameter for quantitative assessment of the manifestation of the effect of the global dynamic chaos. This quantity is the number of periods of $RE$ rotation in the Keplerian orbit required for its departure into the ionization continuum $\langle N\rangle \approx \tau_{diff}\cdot T_{0}^{-1}$.  Here $T_{0} = 2\pi n^{3}$ is the $RE$ rotation time in the Keplerian orbit. The decrease of $\langle N \rangle$ when the Sommerfeld parameter changes $\alpha(l) = 3(l^{2} - 0,25^{2})$ should indicate intensification of the process of chaotic diffusion and vice versa.

\begin{figure}
\begin{center}
\includegraphics[width=0.9\columnwidth,
height=0.7\columnwidth]{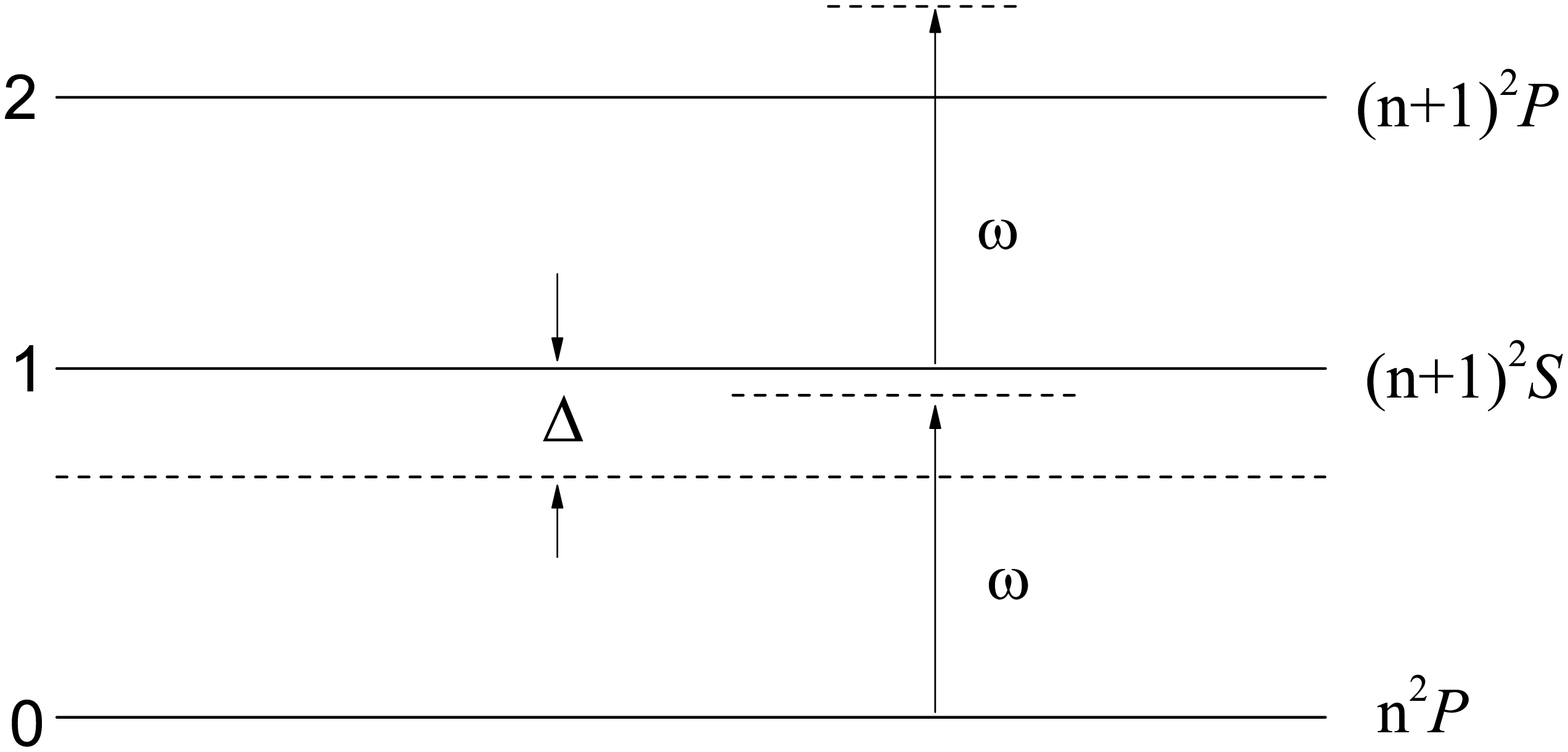}
\end{center}
\caption{Structure of the $RA$ energy levels configuration correspond to the F\"{o}rster resonance case.}
\label{fig:fig5}
\end{figure}

From the standpoint of quantum mechanics, diffusion ionization of $RA$ in the external field can be considered as an analogue of multiphoton ionization.  It is a well-known phenomenon, called the Cooper minimum of photoionization cross-sections, when corresponding values of dipole matrix elements vanish. Under the Seaton criterion \citep{sea83}, this meets the condition of a half-integer value of the difference of the quantum defects (in our case, the difference of $\alpha$) for the two neighboring states involved in the transition. F\"{o}rster resonance is one of manifestations of nonlinear effects in optical transitions between the $RA$ in relatively weak electric fields \citep{bet88}. This resonance occurs when the level of $l$-series is exactly between the two levels of neighboring ($l$-1) or ($l$+1) series (Fig.\ref{fig:fig5}). In particular, such configuration of levels corresponds to the two-photon resonance for the transition in a highly excited alkali atom $\{l+1,n\}\rightarrow\{l,n\}\rightarrow\{l+1,n-1\}$, $\{p,d\}$-series, and $\{l-1,n\}\rightarrow\{l,n\}\rightarrow\{l-1,n-1\}$, $\{s,p\}$-series.

Configuration of highly excited real levels of alkali metals for $\{s,p\}$- and $\{p,d\}$-series is close to the structure of F\"{o}rster resonance, which allows to receive the latter at the values of the electric field intensities of a few V/cm. The F\"{o}rster resonance has received increased attention in literature recently, since it is regarded as a promising tool for the manipulation of an atom in a laser field, and is promising for solving practical problems in quantum information science.

\begin{figure}
\begin{center}
\includegraphics[width=\columnwidth,
height=0.75\columnwidth]{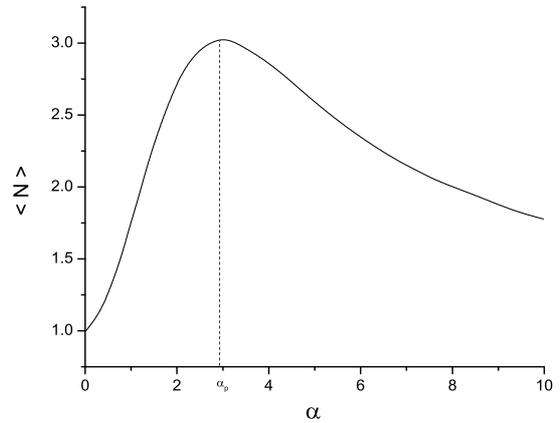}
\end{center}
\caption{Value $<N>$ as Sommerfeld parameter $\alpha$ function. Value $\alpha = 2.81$ correspond to the maximum suppression of the stochastic diffusion.}
\label{fig:fig6}
\end{figure}

Further we will focus on the blocking of dipole matrix elements that are immediately relevant to the issues of the dynamic chaos kinetics. The configuration of levels for the conditions of a F\"{o}rster resonance is similar to the diagram of levels of a three-dimensional quantum oscillator. Its dipole matrix elements are non-zero only for near transitions to adjacent levels, other "long" optical transitions are blocked in $(RA-A)$ quasimolecular complex. It is known that the widths of nonlinear dynamic resonances depend on the values of dipole matrix elements. Blocking of the latter means blocking the dynamic chaos mode (Fig.\ref{fig:fig6}).

The effectiveness of $RE$ random walk across the energy levels in the area of Coulomb condensation can be controlled by external constant electric fields. At a F\"{o}rster resonance this results in repopulation of Rydberg atomic states and of intensification of the light absorption in the infrared range. Note that the model calculations above do not take into account the possibility of $l$-mixing processes. The concentration of Rydberg atoms, and, consequently, the intensity of their corresponding processes of emission and absorption of light as one of the parameters, is determined by the effective lifetime of the excited state $\tau_{eff}$. In the atmospheres of space objects $\tau_{eff}$ value may be related to the collisional diffusion ionization processes discussed above.

\section{Conclusions}

Following from the fact that the F\"{o}ster resonance allows to manipulate
the random walk of $RE$ in the manifold of quantum levels with the redistribution
of the excitation energy of $RA$ and the occurrence of accompanying anomalies
in the IR spectra, in this work it is shown that in the atmospheres of celestial
objects the intensities of corresponding radiative transitions can be associated
with the effects of dynamic chaos via collisional ionization.
There is no doubt that the modern  investigations of Rydberg atoms will lead to
new approaches in atomic physics and its practical applications

\section*{Acknowledgments}
The work was carried out within the EU FP7 Centre of Excellence FOTONIKA-LV and under the partial support by the EU FP7 IRSES Project COLIMA. Also, the authors are thankful to the Ministry of Education, Science and Technological Development of the Republic of Serbia for the support of this work within the projects 176002, III4402.


\bibliographystyle{elsarticle-harv}

\clearpage
\end{document}